\documentclass{elsart}

\usepackage{graphicx}

\renewcommand{\vec}[1]{\mathbf{#1}}

\nocite{*}

\begin{document}

\begin{frontmatter}

\title{Field-asymmetric transverse magnetoresistance
in a nonmagnetic quantum-size structure}

\author[Z]{A.A. Gorbatsevich},
\author[M]{V.V. Kapaev},
\author[M]{Yu.V. Kopaev},
\author[M]{I.V. Kucherenko},
\author[M]{O.E. Omel'yanovskii},
\author[M]{V.I. Tsebro\thanksref{e-mail}}
\address[Z]{Moscow Institute of Electronics, 103498
Moscow, Russia}
\address[M]{P.N. Lebedev Physics Institute,
Russian Academy of Sciences, 117924 Moscow, Russia}
\thanks[e-mail]{Author for correspondence (tsebro@sci.lebedev.ru).}

\begin{abstract}
A new phenomenon is observed experimentally in a heavily doped
asymmetric quantum-size structure in a magnetic field parallel to the
quantum-well layers -- a transverse magnetoresistance which is
asymmetric in the field (there can even be a change in sign) and is
observed in the case that the structure has a built-in lateral electric
field. A model of the effect is proposed. The observed asymmetry of the
magnetoresistance is attributed to an additional current contribution
that arises under nonequilibrium conditions and that is linear in the
gradient of the electrochemical potential and proportional to the
parameter characterizing the asymmetry of the spectrum with respect to
the quasimomentum.
\end{abstract}

\end{frontmatter}

{\bf 1.} An artificially grown asymmetric quantum-size structure in a
magnetic field oriented parallel to the quantum-well layers is a
system with broken fundamental symmetries with respect to inversion of
the coordinates and to time reversal.
These symmetry breakings lead to unusual macroscopic properties.
Specifically, it has been shown theoretically \cite{gorba93,gorba94}
that such a system can possess anomalously large photogalvanic and
magnetoelectric effects. The large values of the photogalvanic effect
were confirmed experimentally in Refs.~3 and 4.

In the present letter we report the observation of a fundamentally new
phenomenon -- a transverse magnetoresistance which is asymmetric with
respect to the sign of the field -- arising in an asymmetric
quantum-size structure. The effect is observed in the case when a
built-in lateral electric field exists in the structure. This usually
happens in a small region near the fused-in metal contact.

{\bf 2.}
Our experimental GaAs/Al$_x$Ga$_{1-x}$As ($x=0.34$) nanostructure is a
heavily doped single i-GaAs quantum well having average width (300\AA)
and bounded on both sides by $\sim$ 300\AA\ wide Al$_x$Ga$_{1-x}$As
barrier layers, uniformly doped with silicon to volume density
c$_{Si}\sim 10^{18}$cm$^{-3}$. The well is separated from the doped
barrier regions by i-Al$_x$Ga$_{1-x}$As spacer layers $\sim$ 100\AA\
wide.

\begin{figure}[h]
\begin{center}
\includegraphics[width=12cm]{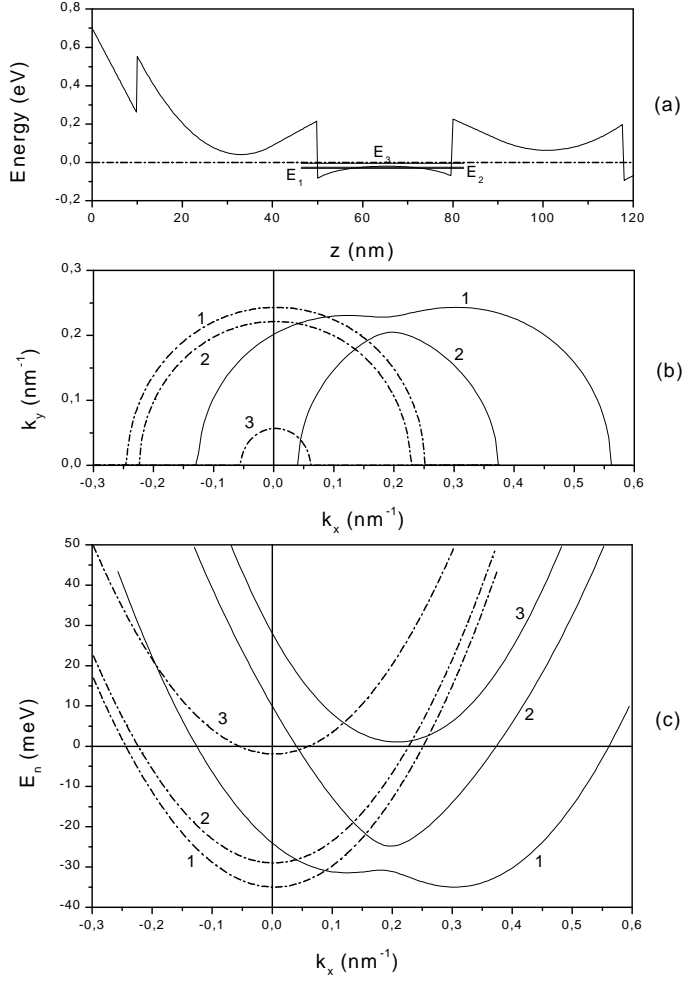}
\caption{Computed potential profile (a) of the conduction band bottom
in the direction of the growth axis of the nanostructure; Fermi
contours (b) and dispersion curves (c) for the three bottom subbands in
1~kOe (dot-and-dash curves) and 70~kOe (solid curves) magnetic fields.}
\label{fig1n}
\end{center}
\end{figure}

A quantum-mechanical calculation of the space-quantization energy
levels in this geometry showed that there are three levels below the
Fermi level $E_F$: $E_1$, $E_2$ and $E_3$, such that
$E_F - E_1 \approx 32$~meV, $E_2 - E_1 = 5\div6$~meV,
and $E_F - E_3 = 1\div5$~meV. The levels $E_1$ and $E_2$ are located
slightly below the convex bottom of the quantum well, so that this
structure can be viewed as a bilayer two-dimensional electronic system
(see Fig.~\ref{fig1n}).

The electronic parameters of the system were determined from
measurements of the Hall effect and Shubnikov-de Haas oscillations with
the magnetic field oriented in a direction normal to the plane of the
nanostructure.
The experimental value of the Fermi energy is
$E_F = (\hbar e/m^{\ast}c) (1/\Delta(1/H)) \approx$ 32~meV
($m^{\ast} = 0.067m_e$), and the two-dimensional
charge-carrier density \linebreak
$n = (e/\pi\hbar c) (1/\Delta(1/H)) \approx 0,9\cdot 10^{12}$cm$^{-2}$
was found to be approximately two times smaller than the carrier
density determined from the Hall constant $R_H$ in weak magnetic fields
$n=1/ecR_H \approx 1,9\cdot 10^{12}$cm$^{-2}$.
These data show that in accordance with the model calculation the
carrier densities in the two bottom subbands are approximately equal,
and the population of the third subband is extremely small because of
the closeness of the bottom of this subband to the Fermi level.

The degree of asymmetry of the nanostructure can be judged according to
the variation of the dispersion curves (Fig.~\ref{fig1n}c) and the
shape of the Fermi contours (Fig.~\ref{fig1n}b) as a function of the
magnetic field for each filled subband (charge-carrier motion is
confined to the $x-y$ plane, and the magnetic field is directed along
the $y$ axis). One can see that despite the very small difference of
the potential energy profile of the nanostructure to the left and right
of the interfaces ($\sim 20$~meV), the magnetic field distorts the
charge-carrier spectrum very strongly, deforming the Fermi contour
along the $x$ axis and leading to a very strong asymmetry of the
dispersion curves $E(k_x)$.

{\bf 3.}
The magnetoresistance measurements were performed by the standard
four-contact method with dc current. The potential contacts were of two
types: a) fused-in metallic (indium) contacts (in this case the
sections of the near-contact region of the nanostructure with a
built-in lateral electric field contribute to the measured electrical
resistance), and b) lithographically prepared lateral contacts through
etched-out extensions of the nanostructure itself (in this case the
near-contact region with the built-in electric field does not make a
contribution).

{\em a) Fused-in indium potential contacts.}

\begin{figure}[h]
\begin{center}
\includegraphics[width=12cm]{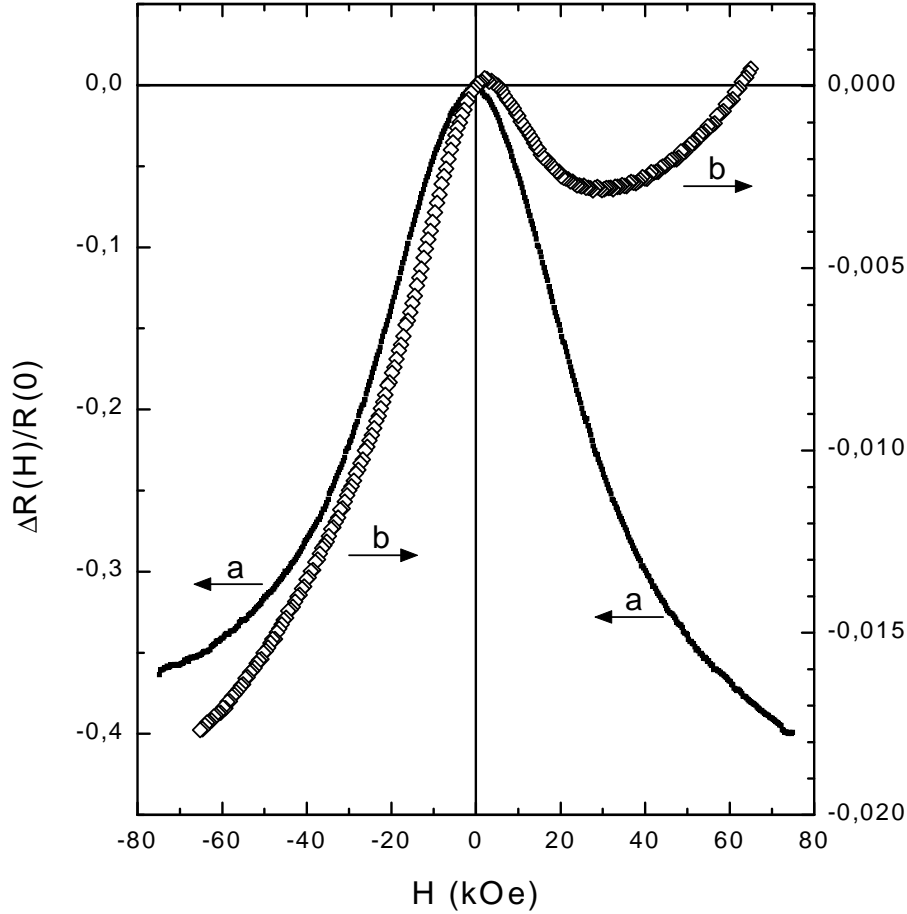}
\caption{Curves of the transverse magnetoresistance at liquid-helium
(a) and room (b) temperatures for opposite orientations of the magnetic
field. The geometry of the contacts is shown in the top portion of the
figure.}
\label{fig2n}
\end{center}
\end{figure}

In this configuration the samples had a rectangular shape with the
dimensions $\sim 2\times 8$~mm and two current contacts fused in along
the entire width of the sample and two $\sim$ 0,5~mm fused-in potential
contacts along one side of the sample, as shown in the upper part of
Fig.~\ref{fig2n}.

Figure 2 shows the measurements of the transverse magnetoresistance at
liquid-helium (curve a) and room (curve b) temperatures for both
directions of the magnetic field. One can see that at liquid-helium
temperature there is a strong negative transverse magnetoresistance
($\Delta R(H)/R(0) \sim - 0.4$ at $H$ = 75~kOe), which differs by
$\sim 10$\% for opposite orientations of the magnetic field, i.e.,
asymmetrically with respect to the direction of H. At room temperature
the magnetoresistance decreases strongly in absolute magnitude to
$\sim 0.01$ in strong magnetic fields, and it becomes asymmetric in H
with respect to not only the magnitude but also the shape of the curves
$\Delta R(H)/R(0)$ (curve b).

It should be emphasized particularly that neither the magnitude nor the
sign of the asymmetry of the transverse magnetoresistance depends on
the direction of the measuring current $J$ through the sample for a
fixed direction of the magnetic field, i.e., they are determined not by
the relative orientation of the vectors $\vec{H}$ and $\vec{J}$
(provided that $\vec{H}\perp\vec{J}$) but by the relative orientation
of the vector $\vec{H}$ and the vector $\vec{l}$ in the direction of
the growth axis ($\vec{H}\perp\vec{l}$).

\begin{figure}[h]
\begin{center}
\includegraphics[width=8cm]{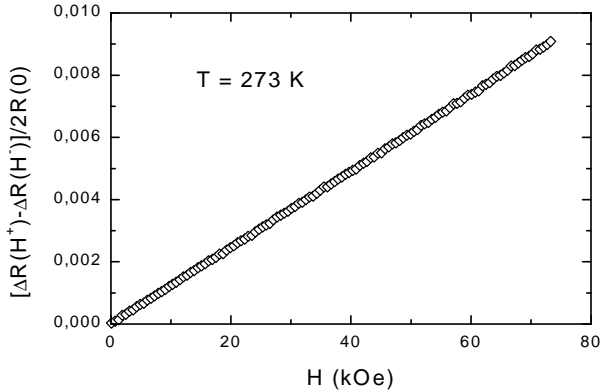}
\caption{Difference between the curves of the room-temperature
transverse magnetoresistance measured for opposite directions of the
magnetic field, plotted versus the absolute magnitude of the field.}
\label{fig3n}
\end{center}
\end{figure}

If the field dependence of the transverse magnetoresistance, measured
for one direction of $\vec{H}$ is subtracted from the corresponding
dependence measured for the opposite direction, then in all cases there
is a strictly linear dependence of the difference obtained on the
absolute magnitude of $H$. This fact is illustrated especially well by
the data obtained at room temperature, where the magnetoresistance is
small and the dependence $\Delta R(H)/R(0)$ has a pronounced
nonmonotonic character (Fig.~\ref{fig3n}).

\begin{figure}[h]
\begin{center}
\includegraphics[width=10cm]{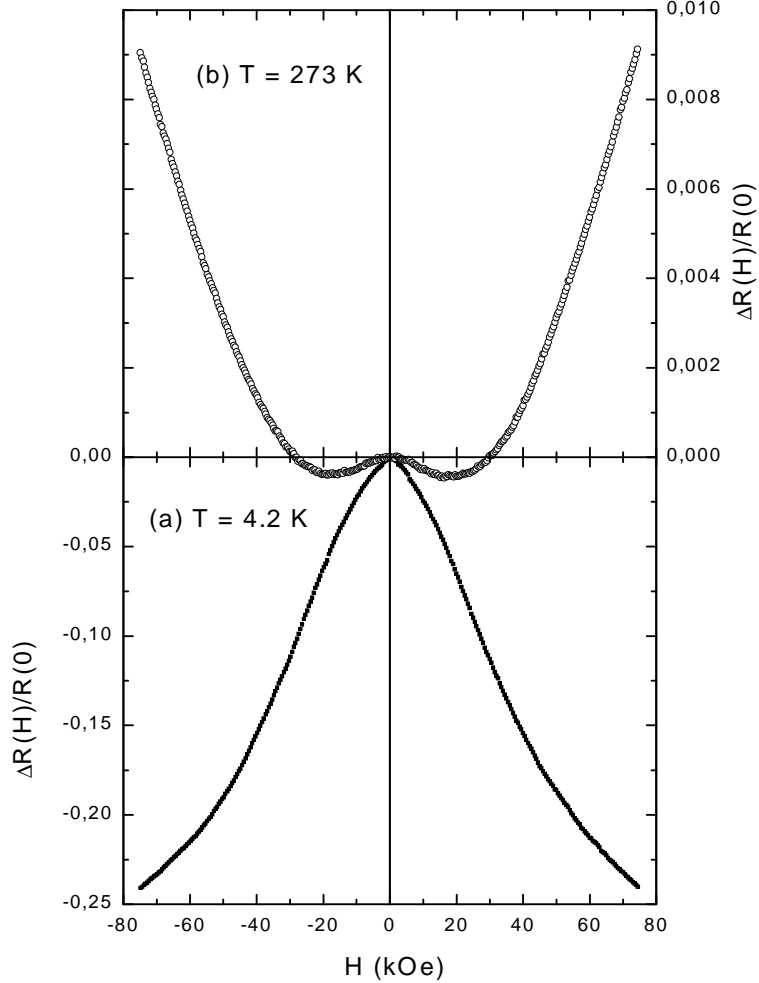}
\caption{Curves of the longitudinal magnetoresistance at liquid-helium
(a) and room (b) temperatures for opposite orientations of the magnetic
field.}
\label{fig4n}
\end{center}
\end{figure}

We note that when the sample is rotated so that the vector $\vec{H}$ is
parallel to the current vector $\vec{J}$ (the case of longitudinal
magnetoresistance), the magnitude, and at high temperatures even the
sign of the magnetoresistance change, but the important fact is that
the asymmetry of the $\Delta R(H)/R(0)$ curves vanishes completely
(Fig.~\ref{fig4n}).

{\em b) Combined fused-in and lithographic potential contacts.}
The region of the nanostructure near the fused-in metallic contact is
a region with the built-in lateral electric field\footnotemark.
This electric field ($\vec{E}_0$), as will be noted below, confers to
this region a nontrivial symmetry, as a result of which there arises a
correction to the conductivity (or current) that is linear in the
magnetic field. Since the observed asymmetry of the magnetoresistance
is proportional to the magnitude and direction of $\vec{E}_0$, it is
obvious that the magnetoresistance asymmetry measured and described
above is a difference effect, which is observable to the extent that
the oppositely directed built-in electric fields $\vec{E}_0^1$ and
$\vec{E}_0^2$ in the near-contact regions of the first and second
potential contacts are unequal. For this reason, it was of interest to
perform measurements of the transverse magnetoresistance on samples
where only one fused-in potential contact is present on one side of the
sample, since in this case an appreciable enhancement of the asymmetry
effect should be expected.

\footnotetext{
Simply depositing indium on the sample surface decrease the surface
barrier (see Fig.~\ref{fig1n}) to $\sim$ 0.5~V. This result in electron
enrichment of the quantum well beneath the contact. On the other hand,
when a metallic indium contact is fused in to some depth, the potential
barrier approaches the quantum well, as a result of which the region
beneath and in direct proximity to the contact becomes depleted of
carriers. This depleted near-contact region possesses a very high
resistivity, and for this reason, despite its small size (estimated
as $\sim 1\div 10$~$\mu m$), its field asymmetric contribution to the
magnetoresistance turns out to be very considerable and determines the
asymmetry of the transverse magnetoresistance of the sample as a whole.
}

\begin{figure}[h]
\begin{center}
\includegraphics[width=10cm]{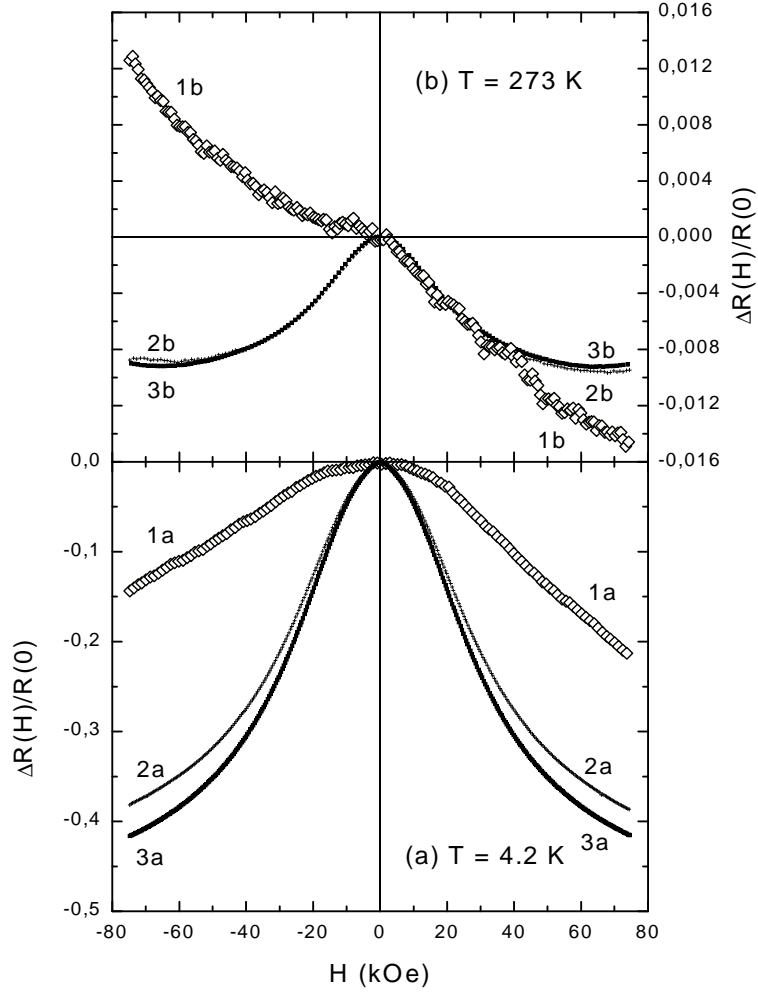}
\caption{Curves of the transverse magnetoresistance at liquid-helium
(a) and room (b) temperatures in the case of combined potential
contacts. Curves la and lb were obtained with potential contacts 1--2;
curves 2a, 2b and 3a, 3b were obtained with contacts 1--3 and 2--3,
respectively.}
\label{fig5n}
\end{center}
\end{figure}

The results of such measurements at liquid-helium and room temperatures
are shown in Fig.~\ref{fig5n}. Three potential contacts were used (see
top part of Fig.~\ref{fig5n}): one fused-in indium contact 1 and two
lateral lithographic contacts 2 and 3, one of which was located close
to the fused-in contact 1 so as to increase appreciably the
contribution of the near-contact region of the fused-in contact 1 to
the total measured magnetoresistance in measurements of the potential
difference with contacts l--2. The distance between contacts 1 and 2
was $\sim$ 0.3~mm and the distance between contacts 2 and 3 was $\sim$
6~mm.

As one can see from the data presented in Fig.~\ref{fig5n}a in the case
of the potential contacts 1--2 the asymmetry of the transverse
magnetoresistance at $T$ = 4.2~K increased appreciably and reached
$\sim$ 50\% (curve 1a). In the case of the potential contacts 1--3 the
asymmetry was $\sim$ 2\%, in accordance with the ratio of the distances
between the contacts 1--2 and 1--3 (curve 2a). Finally, in the case of
the potential contacts 2--3 the magnetoresistance curves are
completely symmetric (curve 3a).

We note that the room temperature magnetoresistance (Fig.~\ref{fig5n}b)
in the case of potential contacts 1--2 does not simply become even more
asymmetric. It even becomes opposite in sign: positive for one
direction of the magnetic field and negative for the other (see curve
lb).

{\bf 4.}
The macroscopic symmetry of the asymmetric system of quantum wells in a
magnetic field parallel to the layers is characterized by a $t$-odd
polar vector \cite{gorba93,gorba94}
\begin{equation} \label{tm}
\vec{T} \propto [\vec{H}\vec{P}],
\end{equation}
where $\vec{H}$ is the external magnetic field, and $\vec{P}$ is a
polar vector characterizing the spatial asymmetry of the system and is
directed perpendicular to the plane of the quantum wells. Physically,
the vector $\vec{T}$ is the toroidal moment density
\cite{ginzb84,gorba89}.

Since $\vec{T}$ and the quasimomentum $\vec{k}$ have the same
transformation properties, the product of $\vec{T}$ by $\vec{k}$ is an
invariant, and the energy spectrum, which can contain all possible
invariants, is asymmetric in the quasimomentum:
$$E(\vec{k}) \ne E(-\vec{k}).$$

It is also known (see Refs.~1 and 2) on the
basis of symmetry considerations that under nonequilibrium conditions
there can exist a macroscopic current
\begin{equation}
\vec{j}=\beta \vec{T},
\label{jbt}
\end{equation}
where the coefficient $\beta $ is due to the departure from
equilibrium. If the source of disequilibrium is photoexcitation, the
above-mentioned anomalously large photogalvanic effect is observed
\cite{alesh93,omel96}.
However, a relation between $\vec{j}$ and $\vec{T}$ similar to
Eq.~(\ref{jbt}) can exist if the nonequilibrium is produced by an
ordinary dissipative (Ohmic) current passed through the system. In this
case, there exists in the system a gradient of the electrochemical
potential, and the dissipative coefficient $\beta$ is linear in the
electric field $\vec{E}$:
$$\beta \propto \alpha (\vec{L}\vec{E}),$$
where $\alpha$ is a scalar and $\vec{L}$ is a polar vector. In the
samples investigated the vector $\vec{L}$ is determined by the built-in
electrostatic field in the near-contact space-charge region. The
expression for the current (\ref{jbt}) in this case can be rewritten in
the form
\begin{equation}  \label{jphle}
\vec{j}=\alpha [\vec{P}\vec{H}](\vec{L}\vec{E})\,.
\end{equation}
In a macroscopically nonuniform system $\vec{E}$ is the gradient of the
electrochemical potential. The current contribution (\ref{jphle})
leads in an obvious manner to an anomalous contribution to the
electrical conductivity, one which is asymmetric with respect to the
magnetic field.

In the microscopic description of the effect, one must substitute into
the general expression for the current density (where $E(\vec{k}$ is
the energy spectrum and $D$  is the $\vec{k}$-space dimension)
\begin{equation} \label{jint}
\vec{j}=\int \frac{\partial E(\vec{k})}{\partial\vec{k}}
f(\vec{k})\frac{d^D\vec{k}}{(2\pi)^D}
\end{equation}
the distribution function $f(\vec{k})$ found from the kinetic equation
\begin{equation} \label{b1}
\left(\vec{v}\frac{\partial}{\partial
\vec{r}}-\frac{e}{\hbar}\left( \vec{E}+\frac{1}{c}\lbrack \vec{v}\vec{H}
\rbrack\right)\frac{\partial}{\partial \vec{k}}\right)
f(\vec{k},\vec{r})=
-\frac{f(\vec{k},\vec{r})-f^0(\vec{k},\vec{r})}{\tau}\,,
\end{equation}

where $\tau$ is the corresponding relaxation time,
$f^0(\vec{k},\vec{r})$ is the equilibrium distribution function,
$\vec{E}=-\nabla \varphi(\vec{r})$ is the electric field,
and $\vec{v} = \partial E(\vec{k})/\partial \vec{k}$ is the velocity of
carriers with the spectrum $E(\vec{k})$ obtained by solving the
Schr\"{o}dinger equation.

For an asymmetric structure, similar to the one investigated in the
present work, the quasiclassical energy spectrum to be substituted into
Eq.~(\ref{jint}) has the form
\begin{equation} \label{ek}
E(k_x,k_y)=E_n(k_x,k_y)+\varphi(x)
=\frac{\hbar^2{k_y}^2}{2m}+E_n(k_x)+\varphi(x)\,,
\end{equation}
where $E_n(k_x)$ is the size-quantized and magnetic-field-quantized
energy spectrum, which in the general case can be easily obtained
numerically (see Fig.~\ref{fig1n}c).

In a system with a built-in potential neither term on the left-hand
side of Eq.~(\ref{b1}) is small, generally speaking, and at equilibrium
they exactly compensate one another. If the deviation from equilibrium
is small, the current
$j=n\mu \nabla F$
(where $F$ is the electrochemical potential) flowing in the system can
be taken as the corresponding small parameter. In this case the kinetic
equation~(\ref{b1}) can be solved by the perturbation method. It can be
shown that the first-order correction to the distribution functions
does not give a contribution to the electrical conductivity that is
asymmetric with respect to the magnetic field. In the case of
nondegenerate charge-carrier statistics, we have for the second-order
correction introduced in the distribution function by the effect under
study
\begin{equation} \label{f2r}
f^{(2)}=-\frac{e^2\tau ^2}{\hbar T}f^{(0)}
\left(\vec{E}_0\frac{\partial}{\partial\vec{k}}\right)
\left(\vec{v}\frac{\partial}{\partial\vec{r}}\right)F\,.
\end{equation}
In this equation $\vec{E}_0=-\nabla\varphi_0$ is the electric field
vector of the built-in field in our system (near the metallic contact).
Upon substitution of expression~(\ref{f2r}) into Eq.~(\ref{jint}), the
corresponding contribution to the current is different from zero only
if the spectrum~(\ref{ek}) is asymmetric with respect to the
quasimomentum. The expression obtained in this manner for the current
has the same form as expression~(\ref{jphle}), where the built-in field
$\vec{E}_0$ enters as the vector $\vec{L}$.

Translated by M. E. Alferieff

\end{document}